\documentclass[onecolumn,secnumarabic,showpacs,preprintnumbers,amsmath,amssymb,nobibnotes,12pt,pra]{revtex4-1}

\usepackage{graphicx}
\usepackage{dcolumn}
\usepackage{bm}

\begin{document}


\title{Classical Light Sources with Tunable Temporal Coherence and Tailored Photon Number Distributions }


\author{Deepak Pandey$^{1,*}$, Nandan Satapathy$^{1}$, Buti Suryabrahmam$^{1}$, J. Solomon Ivan$^{1,2}$ and Hema Ramachandran$^{1}$}
\address{$^{1}$ Raman Research Institute, C.V. Raman Avenue, Sadashivnagar, Bangalore, INDIA-560080\\
$^2$ Presently at Indian Institute of Space Science and Technology, Valiamala, Thiruvananthapuram,
INDIA-695547
\email[$^*$]{deepak@rri.res.in}}
\date{\today}
\begin{abstract}
       We demonstrate a method for the generation of tunable classical light sources with electronic control over its
temporal characteristics and  photon number distribution by modulating coherent light. The  tunability of the  temporal 
coherence is shown through second order correlation ($G^2(\tau)$) measurements 
both in the continuous intensity measurement as well as in the photon counting regimes. 
The generation of desired classical 
photon number distributions is illustrated by creating two light sources - one emitting  thermal
state and the other a specific classical non-Gaussian state. Such tailored light sources with
emission characteristics quite different from that of existing natural light sources are likely to be  
useful in quantum information processing for example in conjunction with photon addition to possibily generate tailored non-clasical states of light.
As a particular application in this direction we outline how  a classical non-Gaussian state generated in this manner 
may be mixed with an appropriate non-classical Gaussian state at a beamsplitter,  to generate non-Gaussian 
entanglement. 
\end{abstract}


\maketitle

\section{introduction}
\label{Introduction}
Understanding light-matter interactions and mechanisms of light generation are of vital importance in  
elucidating many physical processes. While for most processes, a combination of a statistical approach 
along with  Maxwell's classical theory of electromagnetism suffices, several optical 
processes demand a quantum mechanical approach to the electromagnetic field\,\cite{Glauber, 
sudarshan63, Fermi-Quantum}.
The quantum theory of light makes a clear distinction between different sources 
of radiation such as coherent, thermal, and single photon sources\,\cite{Glauber-Coherent}. Radiation 
from the former two sources may  be explained  classically, while light from the latter 
can only be explained quantum mechanically. In other words, the former two states are deemed 
classical, and the   latter non-classical. Due to their quantum nature, non-classical sources of 
light display several counter-intuitive features evoking considerable interest in them.  Non-classical sources of light have been generated experimentally by several techniques\,\cite{single-photon}
and their possible applications in different fields, especially quantum information, are well 
explored\,\cite{Review}. However, recent years have witnessed an emergence of interest in {\it {classical}} 
sources of light \cite{SHIH-THERMAL-IMAGING}-\cite{alessia} 
These have been used in intriguing applications like ghost 
imaging\,\cite{SHIH-THERMAL-IMAGING}-\cite{lugiato06}, and interferometry 
based experiments as in Refs.\,\cite{sam-french, nandan-sam}, and have also formed an ingredient in 
the creation of non-classical states of light\,\cite{zavatta07,parigi09,keisel11}. 

    In this paper we demonstrate a method of creating classical incoherent light sources that can be 
   tailored to mimic light  from a thermal source or can be made to emit light
quite distinct from that emitted by  natural light sources. Utilizing  
the fact that an acousto-optic modulator (AOM) may be  effectively used to  introduce  
phase and intensity fluctuations to light  rapidly and  accurately\,\cite{Elliot,Nandan,quantumwalk},  
we create incoherent light having the 
desired coherence times and intensity statistics, or photon-number distributions, from input coherent light. 
The motivations for creating such sources  are several. This technique offers  an alternative to
the current standard  method of using a rotating  ground glass plate\,\cite{Martien, arecchi65} to generate 
pseudo-thermal light. Further, the electronic control of fluctuations provides a robust and flexible 
procedure for producing tailored classical light sources with predetermined photon emission rates.
Several interesting applications now become possible. For example, it is known that non-classical states can 
be generated by combining
classical light, both coherent and thermal, with single photons (Fock state) in photon addition experiments\,\cite{zavatta07}-\cite{keisel11} ,\cite{gsa}-\cite{Grangier-Science}. The ability to tailor classical states of light to have Gaussian or non-Gaussian   photon 
number distributions  as demonstrated in this paper, widens the field of generation of
non-classical states of light by making many novel forms possible.
 In addition, classical non-Gaussian states with tailored  photon number distributions (PND) may, in 
turn, be used to produce states with tailored non-Gaussian entanglement. This is of importance in the 
quantum information theoretic context, where recent findings suggest non-Gaussian entanglement  to be  
advantageous over Gaussian entanglement\,\cite{solop, dellanno07, dellano08}. 

    This paper is arranged as follows. The experimental setup used in this study is described  in Section 
$2$.  Section 3 describes the methodology for the generation of classical incoherent light with tunable 
temporal coherence. This is achieved by imparting random phase shifts to  coherent light through the 
acousto-optic interaction, in a Mach-Zehnder interferometer (MZI). Measurements of the second-order correlation 
function, $G^2(\tau)$, both using a classical photodetector (intensity-intensity correlation) and photon counting 
detector (photon coincidence detection), show the tunability of temporal characteristics, and the equivalence of the 
two forms of detection. In Section 4, we generate classical light sources with the desired PNDs 
introducing intensity fluctuations to light by suitably modulating the diffraction efficiency of the 
acousto-optic modulator (AOM). Two incoherent states, one  thermal  and the other a classical non-Gaussian 
state were created as illustrative examples. In Section $5$ we discuss  the possible use of 
such tailored classical non-Gaussian state in producing tailored non-classical states and non-Gaussian entanglement. Section 
$6$ summarizes the work presented in this paper.

\section{Experimental Setup}
\begin{figure}[h]
\centering
\includegraphics[width=12.0cm]{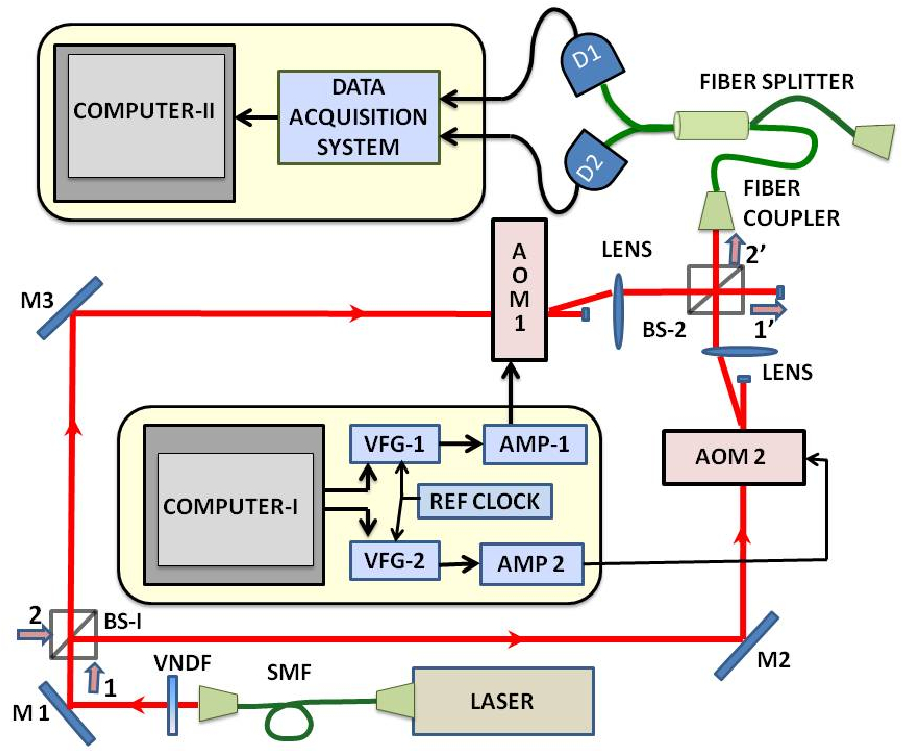}
\caption{Schematic of the  experiment. AMP - amplifier, AOM - acousto-optic modulator, BS - beam splitter, D1,D2 - detectors, M1 - M3 - mirrors, SMF - single mode fiber,  VFG - versatile function generator, VNDF - variable neutral density filter. }
 \label{schematic}
\end{figure} 

The schematic of the experiment is given in Fig. 1. Coherent cw laser light from an external cavity diode laser (Toptica, 767 nm, linewidth $< 5$ MHz) was fiber coupled through a single mode 
polarization maintaining fiber. The light was then passed through a variable neutral density filter (VNDF) ( to control the intensity of the beam) onto 
beam splitter BS1 where it was amplitude divided into two parts. These two beams traversed along the two arms of a Mach-Zehnder interferometer (MZI), which
had an AOM in each of the two arms, arranged such that the first order diffracted light proceeded further on, 
while the undiffracted light was blocked.  After traversing the two arms of the interferometer, the beams were combined 
at BS2, that is, the diffracted light beams of the two arms of the MZI interfered at BS2. Light emerging from one of the exit ports of BS2 was directed into an input of a  50:50 fiber-splitter, and then onto two detectors, D1 and D2. 

 The two AOMs were driven by individual Versatile Function Generators (VFG, Toptica) which operated 
at 80 MHz radio frequency (rf); they were both referenced to a common 10MHz  clock. Using a LabVIEW interface
we could tailor any distribution of phase and intensity fluctuations  in the rf electrical signal  being fed to the AOM on time scales ranging from few hundreds of nanoseconds to seconds. These fluctuations were transferred to the diffracted light by the  acousto-optic interaction thus providing  fine electronic control 
over the phase and intensity of the light.
Measurements at the two detectors were used to determine the  second order correlation 
(intensity-intensity correlation) function. For the case of continuous light intensity measurements, the laser was operated at a power of around $30\,\rm mW$ and  D1  and D2 were two fast photodiodes (Thorlabs PD10A-EC). 
For  the case of photon coincidence detection, the laser light was strongly attenuated, photodiodes D1 and D2 were 
avalanche photo-diode (APD) based single photon counting modules ( SPCM-AQR-15 Perkin Elmer) where an incident 
photon results in a  TTL pulse with a detection efficiency of 65\,\% at 767 nm \cite{spcm}. The outputs of D1 and D2 were 
stored in a PC using data acquisition systems. In the case of classical detectors, a digital storage oscilloscope was 
employed, while for photon counting,  two  counters on a data acquisition card (NI  M-series PCI-6259) were used. 

\section{Generation of Incoherent Light by only phase modulation in an MZI setup}

In this Section, we describe the creation of incoherent sources of light by 
imparting phase jumps to light by means of the AOMs in the MZI (Fig.1). 
The use of a Mach-Zehnder interferometer was motivated by the facts that, i)  phase changes imparted to light 
can be discerned only in an interferometric setup, and ii)  it  enables the creation of   a 
source with intensity fluctuations even though only phase jumps are 
imparted to light. 

\subsection{Theory} 

The action of the MZI in Fig.\ref{schematic} may be mathematically represented 
by the transformation 
matrix $M = B.\Phi.B$ where $B$ ( the operation of a beam 
splitter) and $\Phi$ (the action of the two AOMs) are given by 
\begin{eqnarray}
B =
\frac{1}{\sqrt{2}}
\left(
\begin{array}{cc}
\,\,\,\, 1 & 1 \\
 -1 & 1
\end{array}
\right)
\,\,\,\,  
\rm and 
\,\,\,\,\,\,
\Phi=
\left(
\begin{array}{cc}
 e^{i \phi _1 (t)} & 0 \\
 0 & e^{i \phi _2 (t)}
\end{array}
\right)
\end{eqnarray}
Here $\phi_1(t)$ and $\phi_2(t)$ are the phase shifts imparted to light at AOMs 1 and 2, respectively.
  Thus, in terms of  the  scalar 
wave field, $E_1$, entering one of the ports of $BS1$ (and with no input at its other port) the output at $BS2$ is given by  
\begin{eqnarray}
\left(
\begin{array}{c}
 E_{1'} \\
 E_{2'}
\end{array}
\right)&=& M.
\left(
\begin{array}{c}
 E_1 \\
 0
\end{array}
\right)
\end{eqnarray}
where 1, 2 represent the input ports of BS1 and 1', 2' the output ports of BS2.
It is clear  that the second order intensity self-correlation at the output ports is 
given by
\begin{eqnarray}
G^2_{ii}(\tau) = \frac{\left\langle E_i^\star(t) E_i(t) E_i^\star(t+\tau) 
E_i(t+\tau)\right\rangle_t}{\left\langle E_i^\star(t) E_i(t)\right\rangle_t 
\left\langle 
E_i^\star(t+\tau) E_i(t+\tau)\right\rangle_t}
\label{g2classical}
\end{eqnarray}
where i can take values 1', 2'.  In terms of field operators $G^2_{ii}(\tau)$ is 
given by
\begin{eqnarray} \frac{\left\langle\psi\right|A_i{}^{\dagger}
(t)A_i{}^{\dagger}(t+\tau)A_i(t+\tau)A_i(t)\left|\psi\right\rangle_t}
{\left\langle\psi\right|A_i{}^{\dagger}(t+\tau)A_i(t+\tau)\left|
\psi\right\rangle_t\left\langle\psi\right|A_i{}^{\dagger}(t)A_i(t)\left|
\psi\right\rangle_t}
\label{g2quantum}
\end{eqnarray}
where 
\begin{eqnarray}
\left(
\begin{array}{c}
 \hat{A_{1'}} \\
 \hat{A_{2'}}
\end{array}
\right)= M.
\left(
\begin{array}{c}
 \hat{a_1} \\
 \hat{a_2}
\end{array}
\right),
\end{eqnarray}
with $\left|\psi\right\rangle$ being the input state,  $\hat{a_1}$, $\hat{a_2}$  
the annihilation operators of the input modes, and  $\hat{A}_{1'}, \hat{A}_{2'}$ 
those of the output modes, and $M$ is as defined earlier.  In our setup, $\left|\psi\right\rangle =\left| E_1 \right\rangle 
\otimes \left| 0 \right\rangle$,  where $\left| E_1\right\rangle$ is a  coherent state.
Simplification of Eqs. \ref{g2classical} and \ref{g2quantum}, for the case of complete phase noise, (i.e., uniformly distributed noise, with phase spanning the entire circle)  both yield
\begin{eqnarray}
G^2_{ii}(\tau)  = 1+\left\langle\cos(\delta\phi(t))\cos(\delta\phi(t+\tau))\right\rangle_t
\end{eqnarray}
where $\delta\phi=\phi_1(t)-\phi_2(t)$. Note that we have adjusted the MZI to 
match the dynamical paths in both arms and, therefore, the phase difference is the 
difference in the phases imparted at the two AOMs. 

    The temporal coherence of light, emerging out of the port 2', was determined by intensity correlation 
technique in a standard correlation setup developed by Hanbury-Brown and Twiss \cite {HBT}. Equation\,(6) can 
be easily cast into the form
\begin{eqnarray}
G^2_{2'2'}(\tau)  &=& 1 + 0.5 \xi(\tau),
\label{eqg22}
\end{eqnarray}
where  $\xi(\tau)$ represents the probability of temporal overlap of $\delta\phi(t)$ and $\delta\phi(t+\tau)$. 
In general, for complete phase noise given to AOM 1 and 2 and phase jumps occurring at time intervals $t$  
given by independent distributions $P_1(t)$ and  $P_2(t)$ respectively, we have 
\begin{eqnarray}
\xi(\tau) = \frac{\int_\tau^\infty (t-\tau) P_1(t) d(t)}{\int_0^\infty t P_1(t) d(t)}.\frac{\int_\tau^\infty (t-\tau) P_2(t) d(t)}{\int_0^\infty t P_2(t) d(t)}
\label{xi-eq}
\end{eqnarray}
For the special case where one of the phases, say $\phi_2(t)$, is held constant Eq. \ref{xi-eq} reduces to 
\begin{eqnarray}
\xi(\tau) = \frac{\int_\tau^\infty (t-\tau) P_1(t) d(t)}{\int_0^\infty t P_1(t) d(t)}
\label{xifinal}
\end{eqnarray}

\subsection{Experimental Results}

\begin{figure}[h]
\begin{center}
\includegraphics[width=11.0cm]{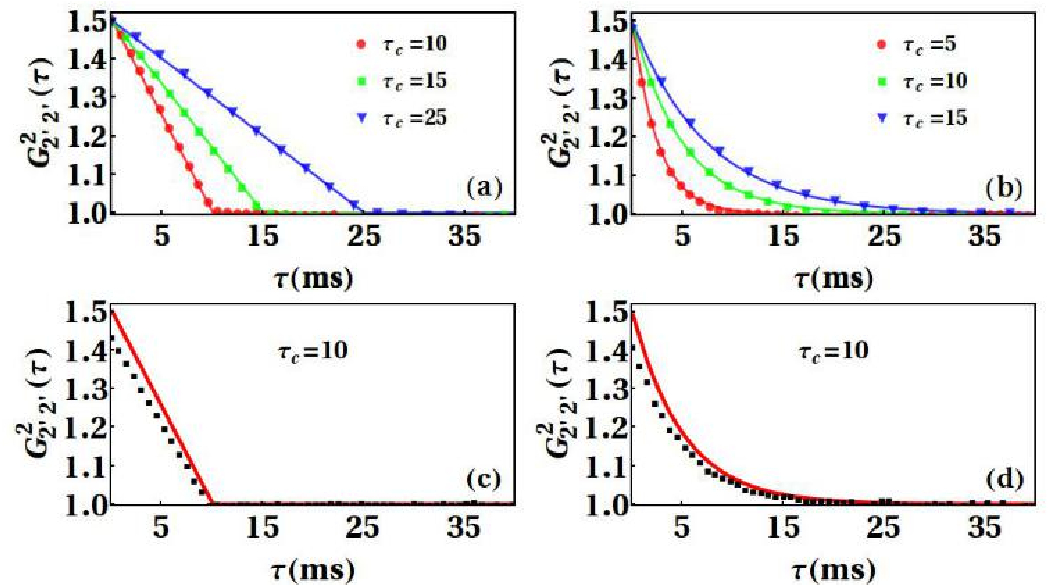}
\end{center}
\caption{Intensity-intensity correlation function,  $G^2_{2'2'}(\tau)$  of the light exiting port 2' of BS2 in Fig.1,  as function of $\tau$ for 
(a) constant dwell time noise to rf of one AOM, and with classical light detection;
(b) exponential distribution of dwell times of  noise in both AOMs, and with classical light detection;
(c) same as (a), but in photon counting regime;
(d) same as (b), but in photon counting regime. }
\label{correl}
\end{figure}

We now present our experimental results. Initially,  random phase jumps were 
imparted to AOM1 only and  at regular intervals, that is,  
$P_1(t)= \delta (t-\tau_c)$. For this case, we see from Eq.\,\ref{xifinal} and Eq.\,\ref{eqg22},

\begin{eqnarray}
G^{2}_{2'2'}(\tau)  &=& 1 + 0.5 \left(1-\frac{\tau}{\tau_c}\right) \quad 
\rm for \quad \tau \leq \tau_c \nonumber \\
                &=&1 \quad \quad \quad \quad \quad \quad \quad \quad \rm for \quad 
\tau \geq \tau_c
\label{eqg22single}
\end{eqnarray}

\noindent that is, on imparting phase noise one expects photon-bunching with a 
 zero-delay second-order 
correlation value 1.5, that falls to 1 for long delays. In the experiment, 
at constant time intervals, $\tau_c$, the rf electrical signal to AOM1 was given 
a random phase jump, distributed uniformly in the interval $(-\pi, \pi)$. 
 The experimentally determined values of $G^2_{2'2'}(\tau)$ are shown as discrete points in 
Fig. \ref{correl}(a)  for three different values of coherence time, $\tau_c$.  The 
continuous curves are $G^2_{2'2'}(\tau)$ as obtained from Eq. \ref{eqg22single}. 

    Next,  AOMs 1 and 2 were driven with  rf voltages with independent, random 
phase jumps  uniformly distributed in the interval  $(-\pi, \pi)$, and at time intervals 
that had independent exponentially falling distributions,  
\quad $P(t)= [\exp({-t/\tau_c})]/\tau_c$, with  $1\,\rm ms$  $\le t \le 100\, \rm ms $ and with a 
mean,  $\sim \tau_c.$ The results for this case are presented in Fig. 
\ref{correl}(b), for three values of $\tau_c$. As 
in the previous case, the agreement between theory and experiment is very good. 

    The above measurements, which utilized  classical detection of intensities with 
photodetectors, were repeated at the  photon counting regime using APD based single-photon counting modules (SPCM). The laser light 
was strongly attenuated, so that, on an average,  every $30\,\mu$s,  there was a 
10\% probability of detecting a photon. Thus, there was, on an average, less than 
one photon in the interferometer at any instant of time. The output pulses from  D1 
and D2 (APD based SPCMs) were fed to two  counters operating at a time bin of  $30\,\mu$s, that was  shorter 
than the mean interval between phase jumps that ranged from $1\,\rm ms$ to $100\,\rm ms$.  
The values of $G^2(\tau)$  obtained from the experiments for the two cases 
(phase jumps to one AOM only at constant intervals and phase jumps to both AOMs with 
exponential distribution of interval between jumps) are shown in Figs. \ref{correl}(c) and 
\ref{correl}(d), along with  theoretically expected values. The agreement is fair. In order  to obtain 
sufficient statistics at single photon detection level, data had to be acquired over 
long durations. Mechanical instability of the interferometer and dark counts  are believed to
have contributed to the deviation from the theoretical curves in the photon counting regime. 

     From the above, it is amply clear that classical light sources exhibiting bunching and with the desired 
temporal coherence characteristics may be created. The results also confirm the 
equivalence between classical intensity-intenstiy correlation and coincidence detection for $G^2(\tau)$ measurement
for classical states of light. 
In the present experiment, the time scales for phase fluctuations are on the milliseconds time scales
due to restriction of the data acquisition card. The setup allows for a temporal variation from 
~$50\,\rm ns$ (currently limited to $500\,\rm ns$ by USB 2.0 communication) to few seconds depending 
upon the stability of the interferometer. This provides an easy way of creating bunched light  with 
long coherence times. Further the bunching can be enhanced and higher values of ${G^2_{2'2'}} (0)$ 
can be obtained by using engineered partial phase noise in this interferometric setup\,\cite{Nandan}.

\section{Tailoring Photon Number Distribution by Intensity Modulation of Light }

In this Section,  we demonstrate the creation of  classical incoherent
states with desired photon number distributions (PND), starting  from an input coherent state. This was achieved by modulating the diffraction
efficiency of a single AOM by the addition of calibrated amplitude noise with the desired characteristics to the input rf electrical signal. 
This, in effect, modulates the transmittivity of the coherent state through the AOM according to the chosen probability 
distribution function, thereby providing a source of classical light with a  tailored photon number distribution.  

\subsection{Theory}

Consider the coherent state $|\alpha \rangle$ being diffracted by an AOM where the transmittivity into the diffracted order
(amplitude of the coherent state) is modulated in time, in the from of ${\cal P}(|\alpha|)$. The modulated coherent state and 
its expansion in terms of number (Fock) states can be written as    
\begin{eqnarray}
\hat{\rho}&=& \int_{0}^{|\alpha_0 |} {\cal P}(|\alpha|) 
|\alpha \rangle \langle  \alpha | d^{2}|\alpha| = \sum_{n=0}^{\infty} p(n) | n \rangle \langle n |,\,\,\,\,{\rm with} \nonumber \\
p(n)&=& \int_{0}^{|\alpha_0 |} d^{2}|\alpha| \, {\cal P}(|\alpha|)\, e^{-|\alpha|^2}\,\frac{|\alpha|^{2n}}{n!}.   
\label{transmit}
\end{eqnarray} 
where 
$p(n)$ is the photon number distribution function and $|\alpha_0 \rangle$ is the diffracted coherent state at maximal transmittivity, and  $\{ |n\rangle\}$ is the Fock basis. 
The ensemble in Eq.\,(\ref{transmit}) is practically realized by appropriately modulating the transmittivity of 
the input coherent state over a sufficient amount of time. The upper limit of $\hat{\rho}$ can be 
taken as infinity if $|\alpha_0|$ is chosen to be much larger than the mean $|\alpha|$.

    We experimentally generate light sources with two  specific PNDs as 
examples. The first is the  thermal state  with an average of $\bar{n}$
photons, namely, $\hat{\rho}_{\rm th}(\bar{n})$\,
\cite{sudarshan63} with
\begin{eqnarray} 
{\cal P}(|\alpha|) \equiv {\cal P}_{\rm th}(|\alpha|)= \frac{1}{ \bar{n}\pi} \exp(-|
\alpha|^2/{\bar{n}} ),
\end{eqnarray}
for which  the probabilities
\begin{eqnarray}
p_{\rm th}(n) = ({1-e^{-\lambda}}) e^{-\lambda n}
\label{eq-thermal}
\,\,\,\,\,\,\, \rm with\,\,\,\,\,\,\, e^{-\lambda}=\bar{n}/(\bar{n}+1),
\end{eqnarray}
 define the PND. The second is the state 
$\hat{\rho}_{\zeta}$
 corresponding to
\begin{eqnarray} 
{\cal P}(|\alpha|) \equiv {\cal P}_{\rm \zeta}(|\alpha|)=  \zeta^2 |\alpha|^2 \exp[- \zeta
  |\alpha|^2],
\end{eqnarray} 
 with an average number of photons $\bar{n}$ and a PND  specified by the probabilities
\begin{eqnarray} 
p_{\zeta}(n)= \left(\frac{\zeta}{\zeta +1} \right)^2\,\frac{n+1}{(\zeta +1)^n}.
\label{eq-non-gaussian}
\,\,\,\,\,\,\, \rm with\,\,\,\,\,\,\, \zeta = \frac{2}{\bar{n}}.
\end{eqnarray}
Note that the state $\hat{\rho}_{\zeta}$ is manifestly non-Gaussian.

The temporal coherence  characteristics of the incoherent light with tailored PNDs, as generated above, is described by the second-order correlation function 
 $G^2(\tau)$,   given by 
\begin{eqnarray}
G^2(\tau)  &=& 1 + (G^2(0)-1)\xi(\tau)
\end{eqnarray}
where $\xi(\tau)$ is as given in Eq.\,(\ref{xifinal}). $G^2(0)=2$ for ${\cal P}_{\rm th}(|
\alpha|)$, $G^2(0)=1.5$ for ${\cal P}_{\zeta}(|\alpha|)$ and 
$G^2(0)=1$ for a coherent state.

\begin{figure}[h]
\centering
\includegraphics[scale=0.5]{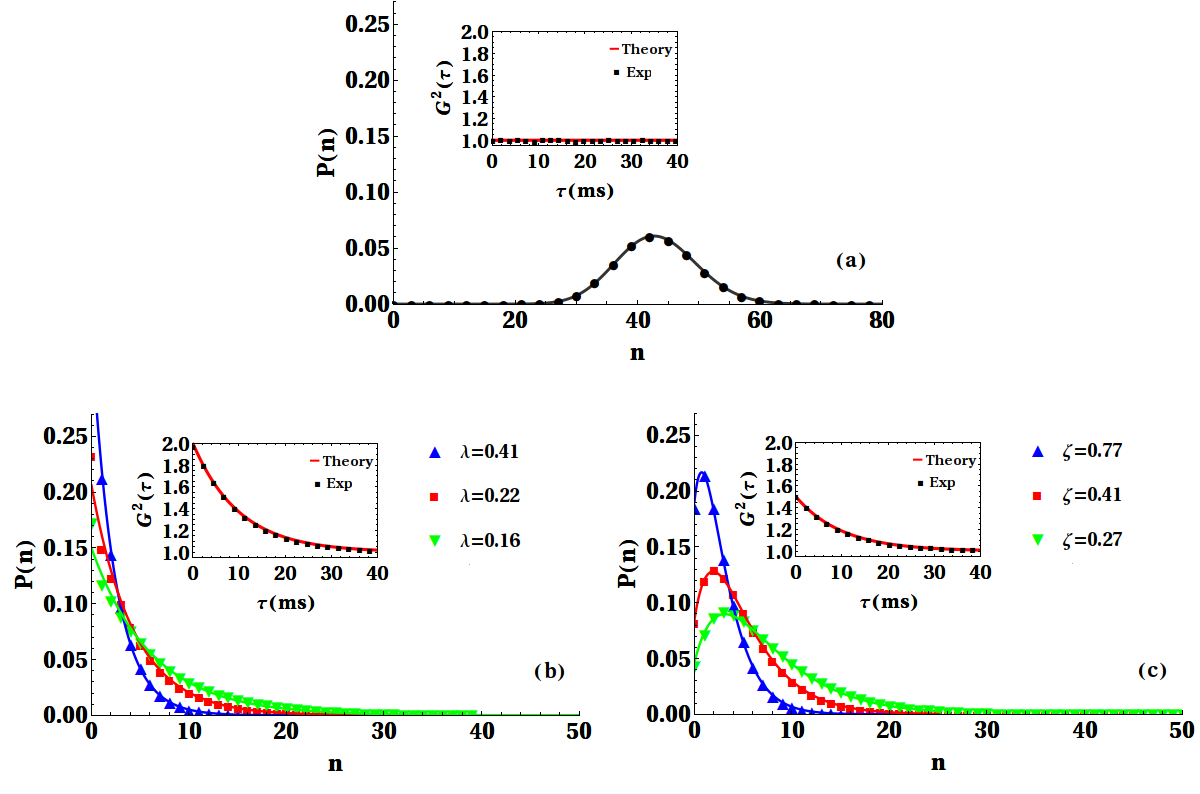} 
\caption{ Photon number distributions, p(n) vs n, as determined from the experiment, with a time bin of 450$\mu$s, for 
(a) maximally diffracted  laser light, in absence of input fluctuations;
(b) light obtained with Gaussian amplitude noise (Eq. 12) applied to the rf, for  different values of $\lambda$;
(c) light obtained with non-Gaussian amplitude noise (Eq.14) applied to the rf,  for different $\zeta$.
Insets show the corresponding $G^2(\tau)$ as function of $\tau$.
\label{histo}
}
\label{fig3}
\end{figure}

\subsection{Experimental Results}

For this part of the experiment  AOM1 of Fig.1 was switched off and  BS2 was 
removed. Thus, only the  light emerging  from AOM2  could reach the detectors, which in this case 
were APD based SPCMs, as described previously. The input laser beam was 
attenuated using neutral density filters to at most 94000 counts per second (cps), suitable for measurement in the photon counting regime
 \cite{spcm2}. The number distribution of 
photons was determined using a time bin of $450\,\mu$s. The size of the time bin, which 
determines the average photon number, was so chosen for the PND measurements to clearly bring out the distinction between coherent and  incoherent states of various parameters. Once chosen this time bin was
fixed for all PND determinations. All measurements lasted for a period of at least 30 minutes to obtain good statistics.

Initially, no noise was added to the rf signal and the maximum rf power was fed to the AOM;  the $+1^{st}$ order diffracted light in this situation constitutes the maximally transmitted  coherent state $|\alpha_0 \rangle$. The PND for this light, as obtained from our measurements, is as shown in Fig. \ref{histo}(a). It gave a Poissonian distribution with a mean photon number of 42 with an estimated average dark counts $2.25\times 10^{-2}$ \cite{spcm,spcm2} per time bin, which is three orders of magnitude smaller than the average photon number in any of our experimental observations. The second-order correlation function, $G^2(\tau)$ for this case is shown in the inset; it is clearly that of a coherent state with $G^2(\tau)=1$ for all $\tau$. This 
was calculated directly from the recorded time series of the counter operating at $30\,\mu$s.
Using the calibration of the AOM transmittivity into the $+1^{st}$ order versus the rf power, and the experimentally determined value of $\bar{n}$ for the maximally 
transmitted coherent state $|\alpha_0 \rangle$, we generate, by appropriate modulation of rf power, light with the desired photon number distribution function. The rf power was fluctuated at random time intervals 
in the range $1$ to $100$ ms with a mean time of $\sim 10$ ms, with the distribution of time intervals 
falling exponentially.

 Then the rf power fed to the AOM was varied such that the 
transmitted coherent state $|\alpha \rangle$ was modulated to 
realise ${\cal P}_{\rm th}(|\alpha|)$ of Eq.\,\,$12$, for three 
different values of $\bar{n}$ ( 1.91, 3.85 and 5.67). The 
emergent light was found to have the PNDs as shown in Fig. 
\ref{histo}(b). These are in good agreement with the 
theoretically expected curves given by Eq.\,13 for ${ p}_{\rm 
th}(n)$. The second-order correlation function for this case is 
shown in the inset. 
 The zero-delay second-order correlation has a value 2, as 
expected for thermal light. This shows 
that temporal modulation of intensity of coherent light leads to the bunching of photons.  
On similar lines, the rf power fed to the AOM was varied such that the  transmitted coherent state $|\alpha\rangle$ was modulated to realise the non-Gaussian function ${\cal P}_{\rm \zeta}(|\alpha|)$ of Eq.\,\,$14$ for three different values 
of $\bar{n}$ (2.60, 4.88 and 7.41). The PNDs
obtained for the emergent light are shown in Fig. \ref{histo}(c), along with the theoretically expected
curves given by Eq.\,\,15 for ${p}_{\rm \zeta}(n)$ . The  
second order correlation function is shown in the inset.  The zero-delay second-order correlation has a value 1.5, as expected for this case. The good agreement between theory 
and experiment for all measurements, both Gaussian and non-Gaussian,  underlines the reliability and efficacy of this method in generating 
tailored classical light sources with desired PND and temporal coherence 
characteristics. Generation of classical non-Gaussian states with desired PND in a 
predetermined manner promises to be  useful in various 
contexts as discussed in the next section.

\section{Application of classical non-Gaussian state in quantum optics}
 As our method offers complete flexibility of providing phase and/or intensity fluctuations with different desired 
probability distributions and on different  time scales, the AOMs can be used to  generate  classical 
incoherent light having properties quite different from existing light sources, opening up new fields for exploration.
 For example it is a simple matter to produce light with temporal incoherence 
but spatial coherence. This is a crucial requirement in  experiments involving photon addition/subtraction to
 incoherent light\cite{parigi09} and  this AOM-based technique is likely to find immediate 
application here. Another feature, likely to prove useful, is the ability  to control the mean photon number with ease, even during the course of a 
measurement, by electronic means.  This has been illustrated in 
Fig. {\ref{histo}(b) and \ref{histo}(c)}, where we have  changed the parameters $\lambda$ and $\xi$ to alter the mean 
photon number. 
\begin{figure}[ht]
\begin{center}$
\begin{array}{c c}
\includegraphics[height=5.0cm, width=7.2cm]{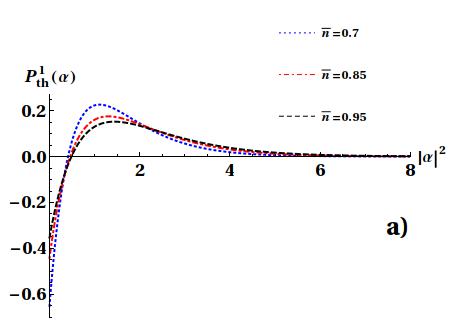}&
\includegraphics[height=5.0cm, width=7.2cm]{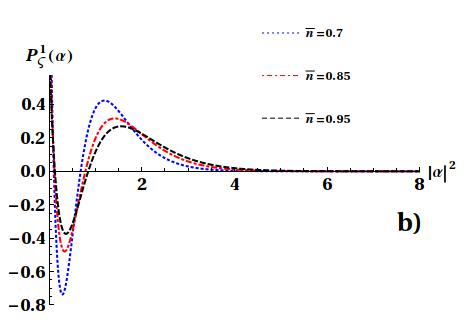}\\
\end{array}$
\caption{The distribution function ${\cal P}(|\alpha|)$ for photon addded thermal state, $\rho^1_{th}$, (Fig.\,\,(4) a)) and photon added classical non-Gaussian state,  $\rho^1_{\zeta}$ (Fig.\,\,(4) b)) given by Eq.\,14. The plots for three different values of average photon number are given for each case. Note the qualitative difference between the two states for same average photon number.} 
\label{phot_add}
\end{center}
\end{figure}
\subsection{Non-classical behaviour of photon added classical non-Gaussian state}
\begin{figure}[h]
\begin{center}
\includegraphics[height=7.5cm, width=10cm]{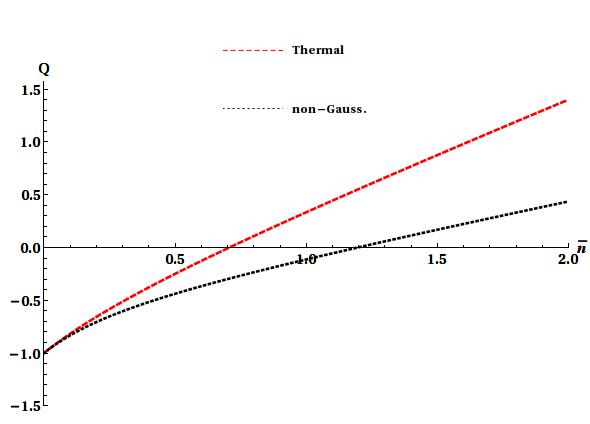}
\caption{The differece in the non-classicality behaviour ($Q < 0$) for the two different photon added states, one with thermal
$\rho^1_{th}$ and other with non-Gaussain state $\rho^1_{\zeta}$. For the same mean photon number $\rho^1_{\zeta}$ shows more negativity compared to state $\rho^1_{th}$. Even more interestingly for mean photon numbers where $\rho^1_{th}$ is classical ($Q > 0$), $\rho^1_{\zeta}$ still continues to be non-classical even upto average photon number $\bar{n} = 1$.
\label{padd3}}
\end{center}
\end{figure}

A significant application of  the tailored classical non-Gaussian  
states generated by this method is in the creation of tailored non-classical non-Gaussian states 
for example when used in conjunction with the photon addition technique. 
  We consider the photon-added thermal state 
$\hat{\rho}_{\rm th}^{1}=\frac{1}{N_1}\,\hat{a}^{\dagger} \,\hat{\rho}_{\rm th}\,\hat{a} $
and the photon-added non-Gaussian state $\hat{\rho}_{\zeta}^{1}=\frac{1}{N_2}\,
\hat{a}^{\dagger}\,\hat{\rho}_{\zeta}\,\hat{a}$. Here the superscript $1$ indicates that we have added one
photon, and $N_1$ and $N_2$ are appropriate normalisations. The 
Glauber-Sudarshan's diagonal weight functions\,\cite{sudarshan63} of the respective 
photon-added states are given by
\begin{eqnarray}
{\cal P}_{\rm th}^{1}(|\alpha|)&=& \frac{1}{\pi \,\bar{n}^3}\,
(|\alpha|^2(1+\bar{n})-\bar{n})\,\exp[-|\alpha|^2/\bar{n}], \, \nonumber \\
{\cal P}_{\zeta}^{1}(|\alpha|)&=&\frac{4}{\pi\,\bar{n}^4\,(1+ \bar{n})}\,
(\bar{n}^2 - 3\bar{n}(\bar{n}+2)|\alpha|^2+ (\bar{n}+2)^2|\alpha|^4) \times \nonumber \\
&& \exp[-2|\alpha|^2/\bar{n}], 
\label{photon-add}
\end{eqnarray}
with $\zeta = \frac{2}{\bar{n}}$ for the latter. As is well known, any state with a 
pointwise non-positive diagonal weight function is nonclassical (quantum). Clearly, both the weight
functions in Eq.\,(\ref{photon-add}) correspond to nonclassical states, and this is made
manifest in Figs.\,(\ref{phot_add}) a) and Figs.\,(\ref{phot_add}) b) . 
Further the nonclassicality (quantumness) 
of the states $\hat{\rho}_{\rm th}^{1}$ and $\hat{\rho}_{\zeta}^{1}$
are qualitatively different. 
To make this observation quantitative, we now evaluate the Mandel parameter $Q$, of 
these states with respect to the mean number of photons. The Mandel parameter $Q$ of a state $\hat{\rho}$ 
is defined as $Q = ({\rm Tr}(\hat{\rho}\,\hat{a}^{\dagger 2}\hat{a}^{2})
- ({\rm Tr}(\hat{\rho}\,\hat{a}^{\dagger}\hat{a}))^2)/
{\rm Tr}(\hat{\rho}\,\hat{a}^{\dagger}\hat{a})$\,\cite{mandel79}. As any state  with 
$Q < 0$ is definitely nonclassical (quantum), it is clear from Fig.\,(\ref{padd3}) that both $\hat{\rho}_{\zeta}^{1}$ and
$\hat{\rho}_{\rm th}^{1}$ show non-classical behaviuor. More interestingly, there are physical situations in which 
$Q$ of $\hat{\rho}_{\zeta}^{1}$ is $<0$ while
$Q$ of $\hat{\rho}_{\rm th}^{1}$ $>0$ for the same mean no. of photons. This suggests 
tailoring the non-Gaussianity of a 
classical state can  have a direct bearing on the quantum features that may emerge when such a state 
is subject to further quantum processing.
\subsection{Generation of non-Gaussian Entanglement}

Here we show how the tailored PNDs of the kind generated above may be used for producing novel forms of entanglement. 

    As  is well known, a beamsplitter preserves non-Gaussianity\,\cite{solo11-1}, and 
non-classicality\,\cite{sudarshan63,aharonov66}. It also generates entanglement 
of input non-classical states\,\cite{knight02}. These features of the beamsplitter may 
be utilized to create non-Gaussian entangled states\,[24] of the form 
\begin{eqnarray}
\hat{\rho}_{\rm out}^{\rm nG}= U_{bs}\, (\hat{\rho}_{\rm G} \otimes 
\hat{\rho}_{\rm
nG} )\,U_{bs}^{\dagger},
\end{eqnarray}
    by appropriately choosing a Gaussian state $\hat{\rho}_{\rm G}$ and a tailored non-Gaussian state
$\hat{\rho}_{\rm nG}$. Here $U_{bs}$ is the beamsplitter unitary.

    Consider  the situation where a single-mode squeezed state $\hat{\rho}_{\rm sq}$ enters one port 
of a 50:50 beamsplitter and a tailored non-Gaussian state 
$\hat{\rho}_{\rm nG}$ enters the other. Note that here we may also tailor the
non-classicality of $\hat{\rho}_{\rm nG}$ through processes such as photon-addition.
Let  $V ={\rm diag}\,\frac{1}{2}(e^{\mu}, e^{-\mu})$ denote the variance 
matrix of the squeezed state with squeezing parameter $\mu$, and  $\hat{\rho}_{\rm nG}$ the tailored non-Gaussian 
state with  average photon number $\bar{n}$.   The resultant state at the output 
of the beamsplitter is definitely entangled when $2 \bar{n} +1 < e^{\mu}$\,
\cite{simon00}, while also remaining  non-Gaussian. Clearly, this a realization of 
a non-Gaussian entangled state. The method is effective even if the
initial non-classicality were to purely reside in $\hat{\rho}_{\rm nG}$.
For instance, say $\mu=0$ ($\hat{\rho}_{\rm G}$ is the ground state), 
and $\hat{\rho}_{\rm nG}$ is a photon-added tailored 
non-Gaussian PND, the resulting $\hat{\rho}^{nG}_{\rm out}$
is definitely non-Gaussian entangled\,\cite{solo11-2}.
In other words, we may choose the initial non-classicality
to reside either in $\hat{\rho}_{\rm G}$ or in $\hat{\rho}_{\rm nG}$, or in both of them,
while simultaneously tuning the non-Gaussianity of $\hat{\rho}_{\rm nG}$,
to generate the desired non-Gaussian entanglement.

\section {Conclusion}
To conclude, we have experimentally demonstrated a method for the  creation of 
tunable classical light using AOMs where the temporal characteristics, coherence 
time,  photon number distribution function and the mean photon number are  electronically 
tuned. Pseudo-thermal light and non-Gaussian classical light 
have been created from coherent laser light, as illustrative examples. Possible applications of such tailored 
light sources, like the generation of tailored non-Gaussian non-classical state as well as tailored non-Gaussian entanglement, have been discussed.
The present proof-of-principle experiments, which display fluctuations on the milliseconds 
timescales, may be easily augmented with currently available technology to higher speeds (nanosecond timescales)
and may be combined with techniques of photon addition, to create novel forms of non-classical light.






\begin{thebibliography}{00}
\bibitem{spcm} {SPCM-AQR-15 is an avalanche photodiode based single photon counting module
with a dead time of 50 ns. The quantum efficiency of the detector is 90 $\%$ at 767 nm. 
The dark count rate is less than 50 counts per second(cps)}.
\bibitem{spcm2} It may be noted that at these count rates the probability of two or more photons generating a single avalanche signal (including the dead time) is negligible, and thus we could reconstruct theoretically expected PNDs for strongly attenuated continuous laser light.
\end{thebibliography}



\end{document}